\documentclass[a4paper]{article}

\usepackage[dvips]{graphicx}

\def\ct{\cite}
\def\be{\begin{equation}}
\def\ee{\end{equation}}

\begin{document}
\date{January 28, 2002}
\title{A causally connected superluminal Warp Drive spacetime\thanks{This work was made possible by through the Advanced Theoretical Propulsion Group (ATPG) Collaboration; current URL: http://www.geocities.com/halgravity/atpg.html}}

\author{F. Loup\footnote{loupwarp@yahoo.com;Lusitania Companhia de Seguros SA, Rua de S Domingos a Lapa 35 1200 Lisboa
Portugal; research independent of employer.}\and R.
Held\footnote{RonaldHeld@aol.com} \and D.
Waite\footnote{FineS137@aol.com}\and E. Halerewicz,
Jr.\footnote{ehj@warpnet.net}\and M.
Stabno\footnote{xcom@kki.net.pl}\and M.
Kuntzman\footnote{MichaelKuntzman@hotmail.com}\and R.
Sims\footnote{tvkwarp5@yahoo.com;University of North Carolina at
Chapel Hill, Chapel Hill, NC 27599  USA}} \maketitle

\begin{abstract}
It will be shown that while horizons do not exist for warp drive
spacetimes traveling at subluminal velocities horizons begin to
develop when a warp drive spacetime reaches luminal velocities.
However it will be shown that the control region of a warp drive
ship lie within the portion of the warped region that is still
causally connected to the ship even at superluminal velocities,
therefore allowing a ship to slow to subluminal velocities.
Further it is shown that the warped regions which are causally
disconnected from a warp ship have no correlation to the ship
velocity.
\end{abstract}

\section{Introduction}
\begin{figure}[p]
\begin{center}
\includegraphics{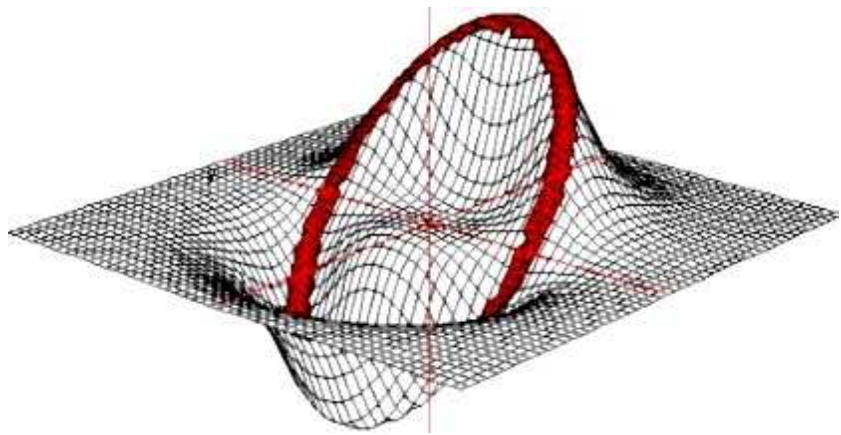}
\end{center}
\caption{Luminal horizon formation. The red region represents where a horizon will form once a warp drive spacetime \ct{lwh01} reaches lumnial velocities.}
\label{horizonform}
\end{figure}

\begin{figure}[p]
\begin{center}
\includegraphics{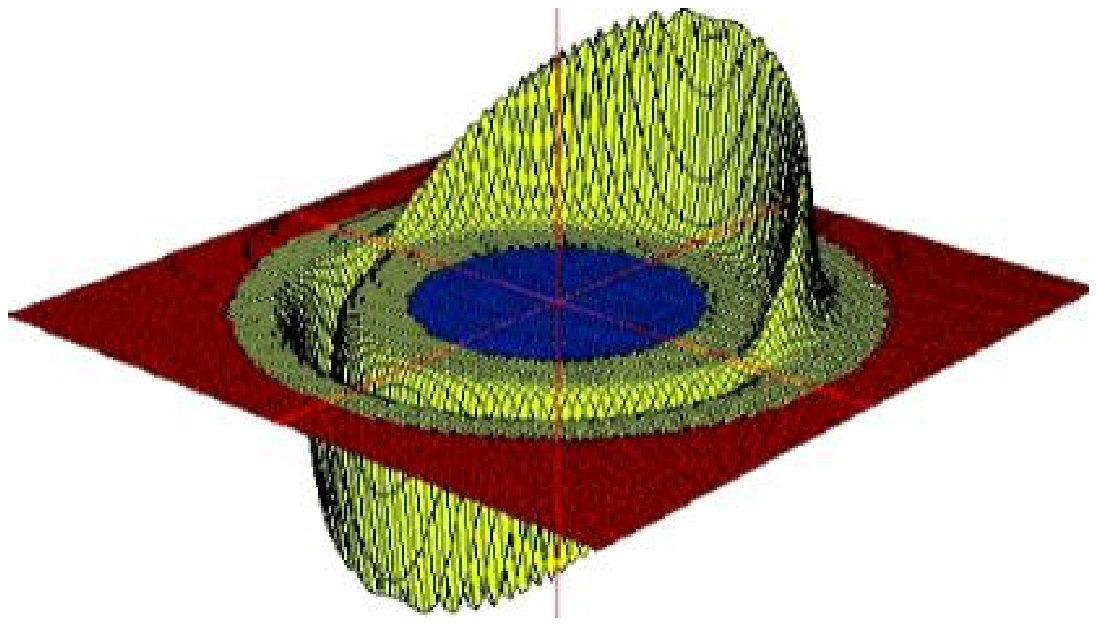}
\end{center}
\caption{Superluminal warp bubble frame regions. The blue region
is the remote frame horizon, the yellow region is the Pfenning
region, and the red region is the ship frame horizon.}
\label{luminalcontrol}
\end{figure}

One of the many obstacles posed by rightfully skeptical physicists
against the warp drive is the appearance of horizons when a ship
travels at superluminal velocities (see figure \ref{horizonform}). This is a problem, to control
the speed of the ship because if the bubble becomes causally
disconnected from the ship then observers in the ship frame cannot
turn off the bubble so the ship cannot reduce its velocity. If the
ship becomes causally disconnected from the bubble then possible
voyages to Messier--42 Orion nebula at 1500 light-years from Earth
or Messier--1 at 6000 light years from Earth would be impossible
because the ship being causally disconnected from the bubble
cannot turn off the bubble and cannot reach its destination.  If a
ship is causally disconnected from the bubble then the warp drive
must be turned off from outside the ship's frame and we don't know
if there is ``someone out there" at Orion or Crab to turn off the
run away warp bubble. In this work we show that while part of the
warped region becomes causally disconnected from a ship when the
ship is luminal or superluminal the behaviour of that part does
not depend on the ship speed and can be engineered while the ship
is still subluminal. Also it shown the control region of the
ship's velocity remains in the portion of the warped region that
is still casually connected to the ship (see figure \ref{luminalcontrol}).

\section{two-dimensional warp drive}

In order to explore the superluminal control problem of the warp
drive we now set up the mathematics which define the physical
horizons (the red region of figure \ref{horizonform}).  In order
to do so it will be necessary to write a two dimensional ESAA
metric \ct{lwh01} written in the Alcubierre formalism:
\be
ds^2=-A^2dt^2+[dx-v_sf(r_s)dt]^2
\ee
where
\[
dx=dx'+v_sdt
\]
\[
v_s={dx_s\over dt}
\]
requiring that
\be
ds^2=-A^2dt^2+[dx'+v_sdt-v_sf(r_s)dt]^2
\ee
where we can replace with
\be
ds^2=-A^2dt^2+[dx'+v_sdt(1-f(r_s))]^2 \ee for simplicity we write
\[
1-f(r_s)={\cal S}(r_s).
\]
So that we arrive at:
\be
ds^2=-A^2dt^2+dx'^2+2v_s{\cal S}(r_s)dx'dt+(v_s{\cal
S}(r_s))^2dt^2 \ee which leads to
\be
ds^2=A^2dt^2-(v_s{\cal S}(r_s))^2dt^2-2v_s{\cal S}(r_s)dx'dt-dx'^2
\ee or
\be
ds^2=[A^2-(v_s{\cal S}(r_s))^2]dt^2 - 2v_s{\cal S}(r_s)dx'dt -
dx'^2 \label{proposed} \ee thereby arriving at a two-dimensional
ESAA spacetime necessary to discuss the mathematics behind the
`horizon problem,' and how to overcome it.

\subsection{Two-dimensional ESAA Hiscock Horizons}

In order to discuss the `horizon problem' we will be improving
upon the paradigm set forth by Hiscock \ct{His}.  Such that an
ESAA-Hiscock ship frame metric can be written from:
\be
ds^2=-H(r_s)dT^2 + {A^2(r_s)dx'^2\over H(r_s)}
\label{horizon1}
\ee
with
\be
dT = dt-{v_s{\cal S}(r_s)\over H(r_s)}dx'
\ee such that we can receive
\begin{eqnarray}
ds^2 = &-H(r_s)dt^2 - 2v_s {\cal S}(r_s)dx'dt/H(r_s)\nonumber \\ &
+(v_s{\cal S}(r_s)/H(r_s))^2 dx'^2+ A^2(r_s)dx'^2/H(r_s)
\end{eqnarray}
by choosing
\be
H(r_s)= A^2-(v_s{\cal S}(r_s))^2
\label{horizon}
\ee
we have a horizon function.  We can thus have the corresponding line element
\begin{eqnarray}
ds^2 = &-H(r_s)dt^2 + 2v_s{\cal S}(r_s)dx'dt - A^2dx'^2/H(r_s)\nonumber \\ &+ H(r_s)dx'^2
/H(r_s) + A^2dx'^2 /H(r_s)
\end{eqnarray}
which reduces to
\be
ds^2=-H(r_s)dt^2 + 2v_s{\cal S}(r_s)dx'dt + dx'^2
\ee
or
\be
ds^2 =- [A^2 - (v_s{\cal S}(r_s))^2]dt^2 + 2v_s{\cal S}(r_s)dx'dt
+ dx'^2 \label{propst} \ee

\section{Pfenning piecewise function in terms of A}

Starting with an arbitrary two-dimensional horizon (\ref{horizon})
we can now begin to define the value of $A$.  In the Pfenning
integration limits $R-(\Delta/2)$ and $R+(\Delta/2)$ \ct{Pfe98},
whereby we set $\Delta=2/\sigma$ and $\sigma=14$ from the
Alcubierre top hat function
\[
f(r_s)={\tanh [\sigma(r_s+R)]-\tanh [\sigma(r_s-R)]\over 2\tanh (\sigma R)}.
\]
By the Pfenning limits the values for the lapse function becomes
\be A(r_s) = \left\{ {\begin{array}{*{20}c}
   1 & {r_s  < R -(\Delta/2)}  \\
   {\kappa _L } & {R - (\Delta/2) \le r_s  \le R + (\Delta/2)}  \\
   1 & {r_s  > R + (\Delta/2)}  \\
\end{array}} \right.
\label{A}
\ee
Where $\kappa_L$ is a large constant, we also note that A \emph{can not} be a function of the
speed.  We now wish to introduce the values of the Pfenning Piecewise function $f(r_s)$.
\be
f(r_s ) = \left\{ {\begin{array}{*{20}c}
   1 & {r_s  < R - (\Delta/2)}  \\
   {1 - (1/A)r_s  - R} & {R - (\Delta/2) \le r_s  \le R +(\Delta/2)}  \\
   0 & {r_s  > R + (\Delta/2)}  \\
\end{array}} \right.
\label{f}
\ee
and now the ESAA ship frame Piecewise function ${\cal S}(r_s)$
\be
{\cal S}(r_s ) = \left\{ {\begin{array}{*{20}c}
   0 & {r_s  < R - (\Delta/2)}  \\
   {(1/A)r_s  + R} & {R-(\Delta/2) \le r_s  \le R +(\Delta/2)}  \\
   1 & {r_s  > R + (\Delta/2)}  \\
\end{array}} \right.
\label{g}
\ee

The Pfenning Piecewise and ESAA Piecewise functions are defined in
function of the term A. This will have some advantages that will
be shown in the work -we will study now the behaviour of the
ESAA-Hiscock Horizon function in three situations:

\begin{itemize}
\item 1-ship subluminal ($v<1$)
\item 2-ship luminal ($v=1$)
\item 3-ship superluminal ($v>1$)
\end{itemize}

We do so by defining the ESAA-Hiscock horizon (\ref{horizon}) with
the following function
\be
H(r_s ) = \left\{ {\begin{array}{*{20}c}
   1 & {r_s  < R - (\Delta/2)} &{H(r_s) > 0}  \\
   {A^2-(v_s/A)^2} & {r_s = R-(\Delta/2)} & {H(r_s ) > 0}  \\
   {A^2} & {r_s = R} & {H(r_s) > 0}  \\
   {A^2-(v_s/A)^2} & { r_s = R+(\Delta/2)} & {H(r_s ) > 0}  \\
   {A^2-{(v_s/A)(r_s-R)}^2} & { R-(\Delta/2) < r_s < R+(\Delta/2) }  \\
\end{array}} \right.
\label{h}
\ee

\subsection{subluminal ship velocities}

It is clear why there are no horizons for the proposed spacetime
(\ref{propst}) with the functions
(\ref{A},\ref{f},\ref{g},\ref{h}). Since in this case one is left
with the general definition $H(r_s )>0$.  Since A is large from the
above expressions it can be seen that the ESAA-Hiscock Horizon
function \emph{never} drops to zero. It is also noted that A is
not function of the speed and A is included in the definition of
the Piecewise continuous functions that warrants for large A the
ship will be \emph{always} connected to the region from $r_s=0$
(ship location) to $r_s=R+(\Delta/2)$ (upper Pfenning limit). Since we have
$A=1$ and ${\cal S}(r_s)=1$ for $r_s>R+(\Delta/2)$ we obtain
\[
H(r_s)=1-v_s^2 H(r_s)>0
\]
telling us that subluminal warp shells are causally connected to
the ship.  In this region A must drop back from a large value at
$r_s=R+(\Delta/2)$ to $A=1$ at $r_s>R+(\Delta/2)$ then part of the
warped region is beyond $R+(\Delta/2)$ since we need exotic matter
to force the A back to 1. Furthermore since A is not function of
the speed if the ship changes its speed the behaviour of A will
not be affected. Since $H(r_s)>0$ the ship is causally connected
to this region and is therefore subluminal.

\subsection{luminal ship velocities}

From the functions (\ref{A},\ref{g},\ref{h}) we can now again set
up the general definition $H(r_s)>0$.  Since $A$ is large from the
above expressions it can be seen that the ESAA-Hiscock Horizon
function \emph{never} drops to zero. Again $A$ is not a function
of the speed and $A$ is included in the definition of the
Piecewise functions that warrants for large A the ship will be
\emph{always} connected to the region from $r_s=0$ (ship location)
to $r_s=R+(\Delta/2)$ (upper Pfenning limit). $H(r_s)=0$ since
$A=1$ and ${\cal S}(r_s)=1$ and $v_s=1$, $r_s>R+(\Delta/2)$, thus
from eq. (\ref{horizon1}), we see that a horizon will appear at
luminal speeds the ship becomes causally disconnected from the
region beyond $R+(\Delta/2)$. Since A is large at
$r_s=R+(\Delta/2)$ and must drop back to 1 when $r_s >
R+(\Delta/2)$ we still need exotic matter beyond $R+(\Delta/2)$ to
drop the value of A back to 1 and this warped region is causally
disconnected from the ship, the ship remains causal until
$r_s=R+(\Delta/2)$.  Providing that A is not function of the speed
then A is not affected when the horizon appears in front of the
ship when the ship gets luminal, the behaviour of A was engineered
when the ship was subluminal. And the part of the speed control
still lies in the region between $\int_{R-
(\Delta/2)}^{R+(\Delta/2)}r_s$ so the ship can change the speed
and go back to subluminal if needed.

\subsection{superluminal ship velocities}

Finally the for the superluminal warp drive we again have the
following definition $H(r_s)> 0$.  Since A is large from the above
expressions it can be seen that the ESAA-Hiscock Horizon function
\emph{never} drops to zero A is not function of the speed and A is
included in the definition of the Piecewise functions that
warrants for large A the ship will be \emph{always} connected to
the region from $r_s=0$ (ship location) to $r_s=R+(\Delta/2)$
(upper Pfenning limit). $H(r_s)<0$ since $A=1$ and ${\cal
S}(r_s)=1$ and $v_s>1$ $r_s>R+(\Delta/2)$, again from eq.
(\ref{horizon1}) we see that a horizon will appear.  At
superluminal speeds the ship becomes causally disconnected from
the region beyond $R+(\Delta/2)$. Assuming a continuous spacetime
(Alcubierre) $H(r_s)>0$ at $r_s=R+(\Delta/2)$ but $H(r_s)<$ 0 at
$r_s>R+(\Delta/2)$, somewhere between $H(r_s)=0$ and then we have
the horizon. Since A is large at $r_s=R+(\Delta/2)$ and must
reduce to 1 when $r_s>R+(\Delta/2)$ we still need exotic matter
beyond $R+(\Delta/2)$ to lower the value of A back to 1 and this
warped region is causally disconnected from the ship which remains
causal until $r_s=R+(\Delta/2)$.  Providing that A is not function
of the speed then A is not affected when the Horizon appears in
front of the ship when the ship goes superluminal. The behaviour
of A was engineered when the ship was subluminal and the part of
the speed control still lies in the region between
$\int_{R-(\Delta/2)}^{R+(\Delta/2)}r_s$ so the ship can change the
speed and go back to subluminal if needed.

\section{energy momentum tensor}
Consider now the following stress energy momentum tensor for a ship frame ESAA-warp metric
\be
T^{00}=-\frac{v_s^2}{4}{1\over 8\pi}\left({d{\cal S}(r_s)\over
dr_s}\right)^2\left({\sigma\over r_s}\right)^2 \frac{1}{A^4} \ee
defining $d{\cal S}(r_s)/dr_s=1/A$, implies that
\be
T^{00}=-{v_s^2\over 4}{1\over 8\pi}\left({\sigma\over
r_s}\right)^2 \frac{1}{A^6} \ee which has the capacity to lower
the negative energy densities of a warp drive spacetime even
further.\footnote{It is also noted that large extreme values for A
can affect the curvature of the spacetime in question, thereby
reducing velocity unless the Pfenning warped regions $R\pm
(\Delta/2)$ are enlarged.}

\section{on colliding warp shells}
The remote frame ESAA warp drive metric is given by:
\be
ds^2=-A^2dt^2 + [dx-v_sf(r_s)dt]^2 \ee with $A=1$ inside and
outside the ship frame and in the warped region comes to some
large value $k_l$.  The function $f(r_s)$ has the ordinary values
for top hat functions except at:
\[
f(r_s)=1-(1/A)(r_s-R) \iff R-(\Delta/2) \leq r_s \leq R+(\Delta/2)
\]
for calculating horizons we are concerned with the region $g_{00}=
[A^2-(v_sf(r_s))^2]$, so we will examine the behaviour of $g_{00}$
with $v_s r_s < R+(\Delta/2)$
\be
g_{00}= 1-v_s^2 \iff v_s < 1 \ee this is causally connected to the
remote frame, while  $v_s \geq 1$ is disconnected from the
remote frame, only with the conditions $v_s<1 \leftrightarrow  g_{00} > 0$ do horizons fail
appear. Another way to remove the horizons is to modify the space such
that
\be
g_{00}= A^2- (v_s[1-(1/A)(r_s+R)])^2 \ee thus providing a large
constant value for A, $g_{00}> 0$ thus this region becomes
causally connected to the remote frame.

As seen from remote observer in flat spacetime $g_{00} = 1$, thus
when the ship is superluminal the horizon lies at
$r_s>R+(\Delta/2)$.  From the ship frame this forms the ship
horizon, the ship is causally connected from the region $r_s=0$ to
$r_s=R+(\Delta/2)$  When the ship is superluminal the horizon lies
at $r_s< R-(\Delta/2)$ for a remote frame observer, this is the
remote frame horizon.  The remote frame is causally connected from
$r_s > R+(\Delta/2)$ at a great distance and remains connected
until $r_s=R-(\Delta/2)$ the ship frame cannot send signals to
$r_s > R+(\Delta/2)$ the remote frame cannot send signals to $r_s
< R-(\Delta/2)$ but the region between
$\int_{R-(\Delta/2)}^{R+(\Delta/2)}r_s$ remains connected to both
observers if an astronaut changes the speed the outer parts of the
bubble will react although the astronaut cannot communicate with
the outer parts of the bubble, thus if the region between
$\int_{R-(\Delta/2)}^{R+(\Delta/2)}r_s$ is common to both
observers the bubble will not collapse.

\subsection{Preprogrammed A}

Although we set up to define A by "pre-programmed exotic matter''
which does not change when the ship pass from subluminal to
superluminal velocity (see figure \ref{luminalcontrol}), we have
not defined $k_l$. For a Pfenning Piecewise behaviour of A in the
upper Pfenning limit $R+(\Delta/2)$, $A$ still have a large value
to keep this part causally connected to the ship according to
ESAA-Hiscock function (\ref{horizon}) by making $H(r_s) > 0$ even
when $v_s >1$. Then we have the following values for $A$ according
to $r_s$ (already seen from eq. (\ref{h})). Providing that $A$ is
not function of the speed the ship will be disconnected after $r_s
> R+(\Delta/2)$ but this does not affect the behaviour of $A$. The
Pfenning piecewise function is an approximation used first by
Pfenning to simplify calculations and we are adopting Pfenning
techniques here.  We know that the continuous form of the top hat
$f(r_s)$ is 1 in the ship and 0 far from it, there exists a open
interval $\int_{0}^{1}f(r_s)$ when the function $f(r_s)$ starts to
decrease from 1 to 0. It is in that region where the exotic matter
resides which is the continuous equivalent of the Pfenning warped
region.

If we define
\be
A={1\over 2}\left(1+\tanh[\sigma(r_s-R)^2]\right)^{-N}
\ee where $-N$ is an arbitrary exponent\footnote{However from a dimensional
point of view $N=R/\Delta$, such that N becomes a measure of
shell thickness.} designed to reduce stress-energy requirements.
We will have a continuous form of $A$ defined in function of the
continuous Alcubierre top hat $f(r_s)$ and $A$ is function of
$r_s,R,\sigma$ and $N$. This expression can make $A$ be 1 in the
ship and far from it while being large in the warped region...the
region where $f(r_s)$ starts to decrease from 1 to 0.

Below there are numerical evaluations (see table \ref{tab1})
showing the behaviour of $A$ reducing to 1 after the warped region
\emph{even if the ship is disconnected} due to function of
distance $r_s$. And therefore ``pushing" the ESAA-Hisccock horizon
to the outermost layers of the warped region making the speed
controllable by the ship because the major part of the warped
region is connected to the ship so the ship can reduce to
subluminal velocities.

\begin{table}[p]
\caption{\label{tab1}Warp shell numerical evaluations.}
\begin{center}
\begin{tabular}{cccccc}
 $r_s$&$R$&$\sigma$&$f(r_s)$ &$S(r_s)$  &$A$\\
\hline
0& 50 & 0.1 & 1 &0
& 1.023 \\

10& 50 & 0.1 & 0.9997 &0.0002
& 1.1825 \\

20& 50 & 0.1 & 0.9997 &0.0023
& 3.4228 \\

30& 50 & 0.1 & 0.9997 &0.0178
& 8031 \\

40& 50 & 0.1 & 0.9976 &0.1191
& $3.0\times 10^{25}$ \\

50& 50 & 0.1 & 0.5 &0.5
&$2\times 10^{75}$ \\

60& 50 & 0.1 & 0.1192 &0.8807
&  $3.9\times 10^{25}$ \\

70& 50 & 0.1 & 0.0179 &0.9820
& $4.16\times 10^{15}$ \\

80& 50 & 0.1 & 0.0024  &0.9975
& 140.5 \\

90& 50 & 0.1 & 0.0003 &0.9976
& 1.9956 \\

100& 50 & 0.1 & $4.5\times 10^{-5}$&0.9999 &1.0950 \\
\end{tabular}
\end{center}
\end{table}

\subsection{remote frame}

We now introduce a Hiscock horizon function for the remote frame
\be
I(r_s)=A^2 - (v_sf(r_s))^2 \ee therefore the line element of
remote frame observer is
\be
ds^2 = -I(r_s)dT^2 + {A^2\over I(r_s)}\;dx^2
\ee
lending
\be
dT^2 = dt^2 + {2v_sf(r_s)dxdt\over I(r_s)} + \left({v_sf(r_s)\over I(r_s)}\right)^2 dx^2
\ee
recalling that $(v_sf(r_s))^2=A^2-I(r_s)$ yields
\be
dT^2 = dt^2 + {2v_sf(r_s)dxdt\over I(r_s)} + {A^2-I(r_s)\over
I(r_s)^2}dx^2 \ee which reduces to
\be
ds^2 = I(rs)dt^2 + 2v_sf(r_s)dxdt-dx^2 \ee from the definition
$I(r_s)=A^2- (v_sf(r_s))^2$ we have the following spacetime:
\be
ds^2 = [A^2 - (v_sf(r_s))^2]dt^2+2v_sf(r_s)dxdt-dx^2
\ee
then we recovered the ESAA remote frame metric from a equivalent ESAA Hiscock Horizon
function for the remote frame. Thus a remote frame observer would be given from
\be
ds^2 =-I(r_s)dT^2 + {A^2\over I(r_s)}\;dx^2
\ee
If $v_s < 1\leftrightarrow  I(r_s) > 0$
then the region is causally connected to ship and remote frame.
However if $v_s = 1\leftrightarrow I(r_s) = 0$ the horizon appears for the remote frame, this
region while connected to the ship frame becomes causally disconnected from the remote frame, if $v_s > 1 \leftrightarrow I(r_s)< 0$ and assuming a continuous spacetime $I(r_s)< 0$ at $r_s < R-(\Delta/2)$ between $R-(\Delta/2) \leq r_s \leq R+(\Delta/2) I(r_s) > 0$ then somewhere between $R-(\Delta/2)$ and $r_s < R-(\Delta/2)$ a horizon appears which is causally disconnected from the remote frame while connected to the ship
frame.

\begin{figure}[ht]
\begin{center}
\includegraphics{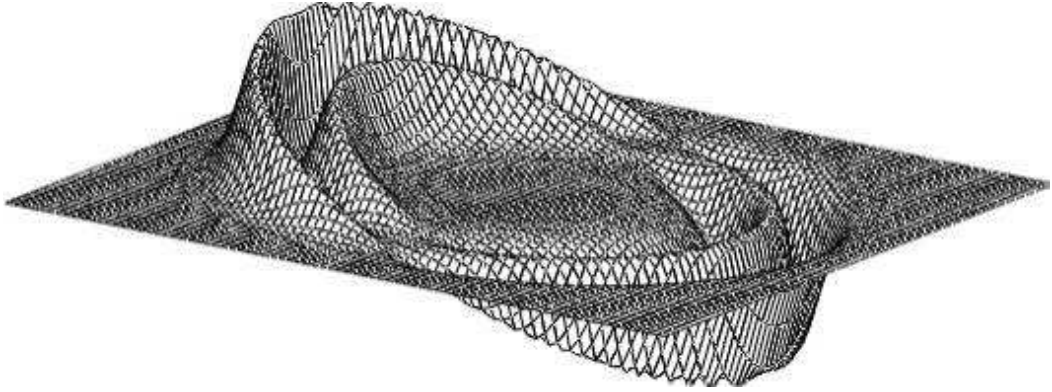}
\end{center}
\caption{Superluminal Warp Drive with control maintained with
inner and outer warp shells defined by
$A=(1+\tanh[\sigma(r_s+R)]^2)/2$, which simulates figure \ref{luminalcontrol}. Graphed with the following
parameters $v_s=1$, $\sigma=2$, and $R=1.8$, $\Delta=0.1$ for the
flat outer Pfenning region.}
\label{2shells}
\end{figure}

If we utilize the top hat function (\ref{f}) for the warped region
$R-(\Delta/2)\leq r_s\leq R+(\Delta/2)$ then one has
\be
I(r_s)=A^2-v({\cal W})
\ee
with
\be
v({\cal W})=\sqrt{1-{v_s^2 (r_s-R)^2\over A(ct,r_s)^2}} \ee and
providing a large $A^2 >> {v_sf(r_s)}^2$ then $I(r_s) > 0$ and
this region will be causally connected to the remote frame.  The
remote frame ``sees" the Pfenning warped region (the part of the
warped region responsible for the speed), thus the remote frame is
causally connected to this region. If an astronaut changes the
speed because the astronaut is causally connected to this region
then the remote frame will ``see" the changing speed.  For
$I(r_s)=1$ this part of the warped region is always connected to
the remote frame as this part of the warped region must make A
drop back to 1 again this region is connected to the remote frame
observer and is disconnected from the ship frame while luminal or
superluminal.

Thus a signal sent by the ship can go to $r_s=R+(\Delta/2)$ and a
signal sent by remote observer can go to $r_s=R-(\Delta/2)$.
Therefore the region between $R-(\Delta/2) \leq r_s \leq
R+(\Delta/2)$ is ``seen" by both observers ship and remote frame.
Since the ship ``sees" $H(r_s)$ the remote "sees" $I(r_s)$
therefore an astronaut can change the ship speed because this
region is connected to the ship  the remote frame "sees" the speed
being changed because this region is connected to the remote
frame.

So the outer part of the bubble ``suffers" when the speed is
changed although the astronaut cannot communicate with the outer
parts of the bubble and the remote observer cannot communicate
with the inner part of the bubble, thus the bubble remains stable
for these regions.

\section{Conclusion}
In this work it was demonstrated how an $A$ defined as a
Pfenning-Piecewise like style function can resolve the
superluminal control problem of the warp drive. It is assumed that
A do not change its behaviour when the ship passes from subluminal
to superluminal, although we do not provide a source for the
nature of A this will done in a future work.  Although if we use
the original continuous expression for A the geometry of the ESAA
warp drive would be the following.  First make the calculations
obey the following format first giving
$(1+\tanh[\sigma(r_s-R)]^2/2)^N$ in the exponent labeled A and the
final form of a is given by Final Form of Coefficient $A = 1/A$ to
produce the following expression
\be
A=\left({1+\tanh[\sigma(r_s-R)]^2\over 2}\right)^{R/\Delta}.
\ee

\section*{Acknowledgements} The Authors of this work would like to
express the most profound and sincere gratitude to Miguel
Alcubierre for his time and patience (especially this one) during
all the phases of development of this work.

\bibliographystyle{unsrt}

\end{document}